\documentclass[preprint,aps,prc,showpacs,nofootinbib]{revtex4}
\usepackage{amsmath}
\usepackage{amssymb}
\usepackage{graphics}
\usepackage{epsfig}

\newcommand\ba{\begin{eqnarray}}
\newcommand\ea{\end{eqnarray}}
\newcommand\be{\begin{equation}}
\newcommand\ee{\end{equation}}
\newcommand\nn{\nonumber}

\begin{document}

\title{Polarization effects in elastic proton-electron scattering}

\author{G. I. Gakh}
\affiliation{\it National Science Centre "Kharkov Institute of Physics and Technology"\\ 61108 Akademicheskaya 1, Kharkov,
Ukraine }

\author{A. Dbeyssi}
\affiliation{
CNRS/IN2P3, Institut de Physique Nucl\'eaire, UMR 8608, 91405 Orsay, France} 

\author{D. Marchand}
\affiliation{
CNRS/IN2P3, Institut de Physique Nucl\'eaire, UMR 8608, 91405 Orsay, France} 

\author{E.~Tomasi-Gustafsson}
\email[E-mail: ]{etomasi@cea.fr}
\altaffiliation{Permanent address: \it CEA,IRFU,SPhN, Saclay, 91191 Gif-sur-Yvette Cedex, France}
\affiliation{
CNRS/IN2P3, Institut de Physique Nucl\'eaire, UMR 8608, 91405 Orsay, France} 

\author{V.~V.~Bytev}
\affiliation{Joint Institute for Nuclear Research, Dubna, Russia}
\begin{abstract}

The experimental observables for the elastic reaction induced by protons scattering from  electrons are calculated in the Born approximation. The differential cross section and polarization observables have been derived assuming one photon exchange. Numerical estimates are given for spin correlation coefficients, polarization transfer coefficients and depolarization coefficients in a wide kinematical range. Specific attention is given to the kinematical conditions; that is, to the specific range of incident energy and transferred momentum.

\end{abstract}

\maketitle

\section{Introduction}

The polarized and unpolarized scattering of electrons by protons has been widely studied, as it is considered the simpler way to access information on proton structure. The expressions which relate the polarization observables to the proton electromagnetic form factors were firstly derived in Ref. \cite{Re68}, whereas the unpolarized cross section was firstly given in Ref. \cite{Ro50} assuming that the interaction occurs through the exchange of a virtual photon. The importance of the information carried by polarization phenomena was stressed long ago, see also Refs. \cite{Ak58,Sc59,Scofield,Dombey}. In all these works, the main attention was devoted to high energies and to the scattering of electrons from protons. In the scattering of proton from electrons at rest (inverse kinematics) approximations such as neglecting the electron mass no longer hold. Liquid hydrogen targets are considered as proton targets, but any reaction with such targets also involves reactions with atomic electrons, which we will assume to be at rest.

A large interest in inverse kinematics (proton projectile on electron target) has been aroused due to two possible applications: the possibility to build beam polarimeters, for high-energy polarized proton beams, in the relativistic heavy-ion collider (RHIC) energy range \cite{Gl97} and the possibility to build polarized antiprotons beams \cite{Ra04}, which would open a wide domain of polarization studies at the GSI facility for Antiproton and Ion Research (FAIR) \cite{FAIR,PAX}. Indeed, assuming $C$-invariance in electromagnetic interactions, the (elastic and inelastic) reactions $p+e^-$ and  $\bar p+e^+$ are strictly equivalent. 

Concerning the polarimetry of high-energy proton beams,  in Ref. \cite{Gl97} analyzing powers corresponding to a polarized proton beam and an electron target were numerically calculated for elastic proton-electron scattering, assuming the one-photon-exchange mechanism and with the dipole approximation for the proton form factors. It was shown that the analyzing powers, as functions of the proton beam energy $E$, reach a maximum for forward scattering at $E=50$ GeV, where the cross section is small. The authors concluded that the concept of such a polarimeter is realistic for longitudinal as well as transverse proton-beam polarization. On the other hand, in that paper, explicit expressions for the analyzing powers were not given.

The possibility of polarizing a proton beam in a storage ring by circulating 
through a polarized hydrogen target was discussed in Refs. \cite{Ra93,Oe09}. Possible explanations of the polarizing mechanisms were published in a number of papers \cite{Ho94,Me94,Mi05}, and more recently in Refs. \cite{Wa09,Mi08}. Motivated by these works, expressions for the helicity amplitudes, depolarization and transfer polarization coefficients have been derived in Refs. \cite{Obrien1,Obrien2}. 

In this work, we derive the cross section and the polarization observables for proton electron elastic scattering, in a relativistic approach assuming the Born approximation. We derive relations connecting kinematical variables in direct and inverse kinematics. Three types of polarization effects are studied: - the spin correlation, due to the polarization of the proton beam and of the electron target, - the polarization transfer from the polarized electron target to the scattered proton, - and the depolarization coefficients which describe the polarization of the scattered proton which depends on the polarization of the proton beam. Numerical estimations of the polarization observables have been performed over a wide range of proton-beam energy and for different values of scattering angle. 

We discuss the properties of the observables for proton--electron elastic scattering  and compare to the recent and ongoing theoretical and experimental work related to the production and the properties of high-energy polarized (anti)proton beams.

\section{General formalism}
\begin{figure}
\mbox{\epsfxsize=8.cm\leavevmode \epsffile{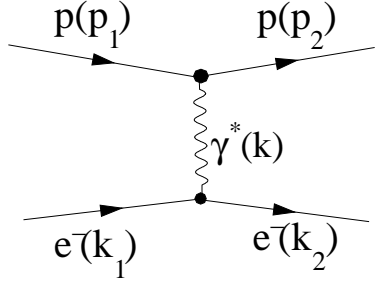}}
\caption{Feynman diagram for the reaction
$p(p_1)+e(k_1)\to p(p_2)+e(k_2)$. The transfer momentum of the virtual photon is $k=k_1-k_2=p_2-p_1$. }
\label{Fig:peFD}
\end{figure}

Let us consider the reaction (Fig. \ref{Fig:peFD})
\be
p(p_1)+e(k_1)\to p(p_2)+e(k_2),
\label{eq:eq1}
\ee
where particle momenta are indicated in parentheses, and $k=k_1-k_2=p_2-p_1$ is the four-momentum of the virtual photon.

A general characteristic of all reactions of elastic and inelastic hadron scattering by atomic electrons (which can be considered at rest) is the small value of the transfer momentum squared, even for relatively large energies of colliding hadrons. Let us first give details of the order of magnitude and the range which is accessible to the kinematic variables, as they are very specific for this reaction, and then derive the spin structure of the matrix element and the unpolarized and polarized observables.
\subsection{Kinematics }
The following formulas can be partly found in Ref. \cite{AR77}. One can show that, for a given energy of the proton beam, the maximum value of the four-momentum transfer squared, in the scattering on the electron at rest, is:
\be
(-k^2)_{max}=\frac{4m^2(E^2-M^2)}{M^2+2mE+m^2}
\label{eq:kmax}
\ee
where m (M) is the electron (proton) mass. Being proportional to the electron mass squared, the four momentum transfer squared is restricted to very small values, where the proton can be considered point-like.
Comparing the expressions for the total energies in two reactions: $s^I=m^2+M^2+2mE$, where $E$ is the proton energy in the elastic proton electron scattering, and $s^D=m^2+M^2+2M\epsilon$, where $\epsilon $ is the electron energy in the electron proton elastic scattering, one finds the following relation between the proton energy and the electron energy, in order to have the same total energy $s^I=s^D$ 
\be
E=\frac{M}{m}\epsilon \sim 2000~\epsilon.
\label{eq:eqE}
\ee
The four momentum transfer squared is expressed as a function of  the energy of the scattered electron, $\epsilon_2$, as:
\be
k^2=(k_1-k_2)^2=2m(m-\epsilon_2),
\label{eq:k2eps2}
\ee
where
\be
\epsilon_2=m\frac{(E+m)^2+(E^2-M^2)\cos^2\theta_e}{(E+m)^2-(E^2-M^2)\cos^2\theta_e},
\label{eq:eqep2}
\ee
and $\theta_e$ is the angle between the proton beam and the scattered electron momenta.

From energy and momentum conservation, one finds the following relation between the angle and the energy of the scattered electron:
\be
\cos\theta_e=\displaystyle\frac{(E+m)(\epsilon_2-m)}
{|\vec p_1|\sqrt{(\epsilon_2^2-m^2)}},
\label{eq:eq3b}
\ee
which shows that $\cos\theta_e\ge 0$ (the electron can never be scattered backward). One can see from Eq. (\ref{eq:eqep2}) that in the inverse kinematics, the available kinematical region is reduced to small values of $\epsilon_2$: 
\be
\epsilon_{2,max}=m\frac{2E(E+m)+m^2-M^2}{M^2+2mE+m^2},
\label{eq:eq3c}
\ee
which is proportional to the electron mass. From momentum conservation, on can find the following relation between the energy and the angle of the scattered proton $E_2$ and $\theta_p$:
\be
E_2^{\pm}=\frac
{(E+m)(M^2+mE)\pm M(E^2-M^2) \cos\theta_p\sqrt{\frac{m^2}{M^2} -\sin^2\theta_p}}
{(E+m)^2- (E^2-M^2) \cos^2\theta_p},
\label{eq:eqE2}
\ee
which shows that for one proton angle there may be two values of the proton energy, (and two corresponding values for the recoi- electron energy and angle-, and for the transferred momentum $k^2$). This is a typical situation when the center of mass velocity is larger than the velocity of the projectile in the center of mass (c.m.), where all the angles are allowed for the recoil electron. The two solutions coincide when the angle between the initial and final hadron takes its maximum value, which is determined by the ratio of the electron and scattered hadron masses, $\sin\theta_{h,max}=m/M$. Hadrons are scattered from atomic electrons at very small angles, and the larger is the hadron mass, the smaller is the available angular range for the scattered hadron.
Let us introduce the invariant 
\be
\nu=k\cdot p_1=E(m-\epsilon_2)+|\vec k_2||\vec p_1|\cos\theta_e=
\frac{k^2}{2m}\left ( E-|\vec p_1|\cos\theta_e\sqrt{1-4\frac{m^2}{k^2}}\right ),
\label{eq:eqvu}
\ee
where ${\vec k_2}$ is the three-momentum of the scattered electron.
One elastic event is represented in the $k^2-\nu$ plane, by two points, which correspond to the intersections of the straight line $k^2+2\nu=0$ with the quadratic expression ($\ref{eq:eqvu}$).

\subsection{Unpolarized cross section}

In the one-photon-exchange approximation, the matrix element ${\cal M}$ of  reaction (\ref{eq:eq1}) can be written as: 
\be
{\cal M}=\frac{e^2}{k^2}j_{\mu}J_{\mu},
\label{eq:eqM}
\ee
where $j_{\mu}(J_{\mu})$ is the leptonic (hadronic) electromagnetic current:
\ba
j_{\mu}&=&\bar u(k_2)\gamma_{\mu} u(k_1),\nn\\
J_{\mu}&=&\bar u(p_2)\left [F_1(k^2)\gamma_{\mu}-\frac{1}{2M} F_2(k^2)\sigma_{\mu\nu}k_{\nu}\right ] u(p_1)\nn\\
&=&  \bar u(p_2)\left [G_M(k^2)\gamma_{\mu}- F_2(k^2)P_{\mu}\right ] u(p_1).
\label{eq:eq2}
\ea
Here $F_1(k^2)$ and $F_2(k^2)$ are the Dirac and Pauli proton electromagnetic form factors (FFs), $G_M(k^2)=F_1(k^2)+F_2(k^2)$ is the Sachs proton magnetic FF, and $P_{\mu}=(p_1+p_2)_{\mu}/(2M)$.
The matrix element squared is:
\be
|{\cal M}|^2=16\pi^2\frac{\alpha^2}{k^4}L_{\mu\nu}W_{\mu\nu},\mbox{~with~}
L_{\mu\nu}=j_{\mu}j_{\nu}^*,~W_{\mu\nu}=J_{\mu}J_{\nu}^*, 
\label{eq:eqma}
\ee
where $\alpha=1/137$ is the electromagnetic fine structure constant.
The leptonic tensor, $L_{\mu\nu}^{(0)}$, for unpolarized initial and final electrons (averaging over the initial electron spin) has the form:
\be
L_{\mu\nu}^{(0)}=k^2g_{\mu\nu}+2(k_{1\mu}k_{2\nu}+k_{1\nu}k_{2\mu}).
\label{eq:eqL10}
\ee
The contribution to the electron tensor corresponding to a polarized electron target is
\be
L_{\mu\nu}^{(p)}=2im\epsilon_{\mu\nu\alpha\beta}k_{\alpha}S_{\beta},
\label{eq:eqLp}
\ee
where $S_{\beta}$ is the initial electron polarization four-vector.

The hadronic tensor, $W_{\mu\nu}^{(0)}$, for unpolarized initial and final protons can be written in the standard form, through two unpolarized structure functions:
\be
W_{\mu\nu}^{(0)}=\left ( -g_{\mu\nu}+\frac{k_{\mu}k_{\nu}}{k^2}\right ) W_1(k^2)+P_{\mu}P_{\nu} W_2(k^2).
\label{eq:eqW}
\ee
Averaging over the initial proton spin, the structure functions $W_i$, $i=1,2$, can be expressed in terms of the nucleon electromagnetic FFs as:
\ba
W_1(k^2)&=&-k^2G_M^2(k^2),\nn\\
W_2(k^2)&=&4M^2\frac{G_E^2(k^2)+\tau G_M^2(k^2)}{1+\tau},
\label{eq:eqWi}
\ea
where $G_E(k^2)=F_1(k^2)- \tau F_2(k^2)$ is the proton electric FF and $\tau=-k^2/(4M^2)$.

The differential cross section is related to the matrix element squared (\ref{eq:eqma}) by 
\be
d\sigma=\displaystyle\frac{(2\pi)^4\overline{\left|{\cal M}\right |^2}}{4\sqrt{(k_1\cdot p_1)^2-m^2M^2}}
\displaystyle\frac{d^3\vec k_2}{(2\pi)^3 2\epsilon_2}
\displaystyle\frac{d^3\vec p_2}{(2\pi)^3 2E_2}
\delta^4(k_1+p_1-k_2-p_2), 
\label{eq:csma}
\ee
where $p_2(E_2)$ is the momentum (energy) of the final proton.
From this point on, formulas will differ from the elastic electron-proton scattering, because we introduce a reference system where the electron is at rest. In this system, 
the differential cross section can be written as 
\be
\frac{d\sigma}{d\epsilon_2}=\frac{1}{32\pi}\frac{\overline{\left|{\cal M}\right |^2}}{m\vec p^2},
\label{eq:eqds}
\ee
where $\vec p$ is the momentum of the proton beam. The average over the spins of the initial particles has been included in the leptonic and hadronic tensors. Using Eq. (\ref{eq:k2eps2}) one can write
\be
\frac{d\sigma}{dk^2}=\frac{1}{64\pi}\frac{\overline{\left|{\cal M}\right |^2}}{m^2\vec p^2}.
\label{eq:eqS}
\ee
The differential cross section over the solid angle can be written as:
\be
\frac{d\sigma}{d\Omega_e}=\frac{1}{32\pi^2}\frac{1}{mp}\frac{\vec k^3_2}{-k^2}\frac{\overline{\left|{\cal M}\right |^2}}{E+m}, 
\label{eq:eqS0}
\ee
where $d\Omega_e= 2\pi d\cos\theta_e$ (due to azimuthal symmetry). We used the relation
\be
d\epsilon_2=\displaystyle\frac{p}{E+m}\displaystyle\frac{\vec k_2^3}{m(\epsilon_2-m)}\displaystyle\frac{d\Omega_e}{2\pi}.
\label{eq:eqDeps}
\ee
The expression of the differential cross section for unpolarized proton-electron scattering, in the coordinate system where the electron is at rest can be written as:
\be
\frac{d\sigma}{dk^2}=\frac{\pi\alpha^2}{2m^2\vec p^2}\frac{\cal D}{k^4},
\label{eq:eqSk}
\ee
 with
\be 
{\cal D}=k^2(k^2+2m^2)G_M^2(k^2)+2\left [k^2M^2+
2mE\left(2mE+k^2\right )\right ]\left [ F_1^2(k^2)+\tau F_2^2(k^2)\right ].
\label{eq:eqD1}
\ee
It  can be written in terms of the Sachs FFs as:
\be 
{\cal D}=k^2(k^2+2m^2)G_M^2(k^2)+2\left [k^2M^2+\frac{1}{1+\tau}\left(2mE+\frac{k^2}{2}\right )^2\right ]\left [ G_E^2(k^2)+\tau G_M^2(k^2)\right ].
\label{eq:eqD2}
\ee
This expression is consistent with Ref. \cite{Gl97}. The differential cross section diverges as $k^4$ when $k^2\to 0$. This is a wellknown result, which is a consequence of the one photon exchange mechanism. 

\section{Polarization observables}
Let us focus here on three types of polarization observables, for elastic proton-electron scattering
\begin{enumerate}

\item The polarization transfer coefficients which describe the polarization transfer from the polarized electron target to the scattered proton, $ p+\vec e\to \vec p+e$.
\item The spin correlation coefficients when both initial particles have  arbitrary polarization, $\vec p+\vec e\to p+e$.
\item The depolarization coefficients which define the dependence of the scattered proton polarization on the polarization of the proton beam, $\vec p+e\to \vec p+e$. In our knowledge, this case was not previously considered in the literature.
\end{enumerate}
The first case is the object of a number of recent papers \cite{Ra04} in connection with the possibility to polarize proton (antiproton) beams. The second case was considered in Ref. \cite{Gl97}, in view of using polarized proton-electron scattering to measure the longitudinal and transverse polarizations of high-energy proton beams. 

Let us calculate the hadronic tensor, when the initial or final proton is polarized. The contribution of proton polarization to the hadronic tensor is
\be
W_{\mu\nu}(\eta_j)=-2iG_M(k^2)\left [MG_M(k^2)\epsilon_{\mu\nu\alpha\beta}k_{\alpha}\eta_{j\beta}+F_2(k^2)(P_{\mu}
\epsilon_{\nu\alpha\beta\gamma}-P_{\nu}\epsilon_{\mu\alpha\beta\gamma})p_{1\alpha}p_{2\beta}\eta_{j\gamma}\right ],
\label{eq:eqWp}
\ee
where the four-vector $\eta_j$ ($j=1,2$) stands for the initial (final) proton polarization.
One can see that all the correlation coefficients in $\vec p\vec e$ collisions are proportional to the proton magnetic FF. This is a well known fact for $\vec e\vec p$ scattering \cite{AR77}. The dependence of the different polarization observables, namely, the spin correlation (the polarization transfer) coefficients on the polarization four-vector of the initial (scattered) proton is completely determined by the spin-dependent part of the hadronic tensor $W_{\mu\nu}(\eta_j)$, $j=1$ $(j=2)$.

\subsection{Polarization transfer coefficients, $T_{ij}$, in the $p+\vec e\to \vec p+e$ reaction}
These polarization observables  describe polarization transfer from the polarized target to the ejectile. The transfer coefficients are also called $T_{i00j}$ in the notations from Ref. \cite{By76}. Here the four subscripts denote, respectively, ejectile, recoil, projectile, and target. The indices $i$, $j$ correspond to $n$, $t$, $\ell$, according to the direction of the polarization vectors of each particle.

The dependence of the scattered proton polarization on the polarization state of the initial electron is obtained by contraction of the spin-dependent leptonic tensor $L_{\mu\nu}^{(p)}$ [Eq. (\ref{eq:eqL10})], and the spin-dependent hadronic tensor $W_{\mu\nu}{(\eta_2)}$ [Eq. (\ref{eq:eqWp})]. The following formula hold in any reference system and can be used to obtain the polarization transfer coefficients:
\be 
{\cal D}T(S,\eta_2)={4mM}G_M(k^2)\left [G_E(k^2)(k\cdot S k\cdot \eta_2-k^2S\cdot \eta_2)
-k^2F_2(k^2)P\cdot S P\cdot\eta_2\right ].
\label{eq:eqpT}
\ee
In the frame where the initial electron is at rest, the polarization four vectors of the electron $S_{\mu}$ and of the scattered proton $\eta_{2\mu}$ have the following components:
\be 
S\equiv(0,\vec\xi),~\eta_2\equiv \left (\frac{1}{M}\vec p_2\cdot\vec S_2,\vec S_2+\frac{\vec p_2(\vec p_2\cdot\vec S_2)}{M(E_2+M)}\right ),
\label{eq:eqpol1}
\ee
where $\vec\xi$ and $\vec S_2$ are the unit three-vectors of the initial electron and scattered proton polarizations in their rest systems, respectively. In the laboratory system (inverse kinematics) one can write $\vec p=\vec k_2+\vec p_2$ and $m+E=E_2+\epsilon_2$.

Using the $P$-invariance of the hadron electromagnetic interaction, one can parametrize the dependence of the differential cross section on the polarizations of the electron target and of the scattered proton as follows:
\be
\frac{d\sigma}{dk^2}(\vec\xi,\vec S_2)= \left (\frac{d\sigma}{dk^2}\right )_{un}\left [1+T_{\ell\ell} \xi_{\ell} S_{2\ell}+ T_{nn} \xi_{n} S_{2n}+ 
T_{tt} \xi_t S_{2t}+T_{\ell t} \xi_{\ell} S_{2t}+ T_{t\ell}\xi_{t} S_{2\ell}\right ],
\label{eq:eqTs}
\ee
where $T_{ik}$, $i,k=\ell,t,n$ are the corresponding polarization transfer coefficients, with the following notations: $\ell$ is the component of the polarization vector along the momentum of the initial proton, $n$ is the component which is orthogonal to the momenta of the initial proton and of the scattered electron (i.e., orthogonal to the scattering plane), and $t$ is the component which is orthogonal to the initial proton momentum and lies in the scattering plane.

At high energy, the polarization transfer coefficients depend essentially on the direction of the scattered proton polarization. Let us choose an orthogonal system with the $z$ axis directed along $\vec p$, $\vec k_2$ lies in the $xz$ plane ($\theta_e$ is the angle between the initial proton and the final electron momenta) and the $y$ axis is directed along the vector $\vec p\times \vec k_2$. Therefore, in this system $\ell\parallel z$, $t\parallel x$ and $n\parallel y$. The explicit expressions for the polarization transfer coefficients are given in Appendix \ref{Poltra}.

\subsection{Polarization correlation coefficients, $C_{ij}$, in the $\vec  p+\vec e \to p+e $ reaction} 
In the reaction involving a polarized proton beam and a polarized electron target, one can derive explicit expressions for the spin correlation coefficients. These coefficients are also called double analyzing powers and denoted $A_{00ij}$ in the notations from Ref. \cite{By76}. 

The contraction of the spin-dependent leptonic $ L_{\mu\nu}^{(p)}$ and hadronic 
$ W_{\mu\nu}(\eta_1)$ tensors, in an arbitrary reference frame, gives:
\be
{\cal D}C(S,\eta_1)= 8mM G_M(k^2)\left [(k\cdot S k\cdot \eta_1-k^2S\cdot\eta_1) G_E(k^2) +
\tau k\cdot \eta_1(k\cdot S+2p_1\cdot S)F_2(k^2)\right ].
\ee
All spin correlation coefficients for the elastic $\vec p\vec e$ collisions can be obtained from this expression and are, therefore, proportional to the proton magnetic FF.

In the considered frame, where the target electron is at rest, the four-vector of the proton beam polarization has the following components
\be
\eta_{1}=\left ( \frac{\vec p\cdot \vec S_1}{M}, 
\vec S_1+  \frac{\vec p(\vec p\cdot \vec S_1)}{M(E+M)} \right ),
\label{eq:eqt1}
\ee
where $\vec S_1$ is the unit vector describing the  polarization of the initial proton in its rest system.

Applying the P-invariance of the hadron electromagnetic interaction, one can write the following expression for the dependence of the differential cross section on the polarization of the initial particles:
\be
\frac{d\sigma}{dk^2}(\vec\xi,\vec S_1)= \left (\frac{d\sigma}{dk^2}\right )_{un}\left [1+C_{\ell\ell} \xi_{\ell} S_{1\ell}+ C_{tt} \xi_{t} S_{1t}+
C_{nn} \xi_n S_{1n}+C_{\ell t} \xi_{\ell} S_{1t}+ C_{t\ell}\xi_{t} S_{1\ell}\right ],
\label{eq:eqpol2}
\ee
where $C_{ik}$, $i,k=\ell,t,n$ are the corresponding spin correlation coefficients which characterize $\vec p\vec e$ scattering.
Here also one expects a large sensitivity of these observables to the direction of the proton beam polarization. Small coefficients (in absolute value) are expected  for the transverse component of the beam polarization at high energies. This can be seen from the expression of the components of the proton-beam-polarization four-vector at large energies, $E\gg M$:
\be
\eta_{1\mu}=(0,\vec S_{1t}) + S_{1\ell}\left (\frac{p}{M},\frac{\vec p}{M} \frac{E}{p}\right )\sim S_{1\ell} \frac{ p_{1\mu}}{M}.
\label{eq:eqpa}
\ee
The effect of the transverse beam polarization appears to be smaller by a factor $1/\gamma$, $\gamma=E/M \gg 1$. This is a consequence of the relativistic description of the spin of the fermion at large energies.

The explicit expressions of the spin correlation coefficients are given in Appendix \ref{Polcor}. One can see that $C_{nn}=T_{nn}$.

\subsection{Depolarization coefficients, $D_{ij}$, in the $\vec  p+ e \to \vec p+e $ reaction} 

In this section explicit expressions for the depolarization coefficients, (also denoted $D_{i0j0}$ in the notation from Ref. \cite{By76}), which define the polarization transfer from initial to final proton, are derived for the reaction $\vec  p+ e \to \vec p+e $. 

The part of the hadronic tensor $ W_{\mu\nu}(\eta_1,\eta_2)$, which corresponds to polarized protons in initial and final states can be written as
\be
W_{\mu\nu}(\eta_1,\eta_2)= A_1\widetilde{g}_{\mu\nu}+A_2P_{\mu}P_{\nu}+
A_3(\widetilde{\eta}_{1\mu}\widetilde{\eta}_{2\nu}+
\widetilde{\eta}_{1\nu}\widetilde{\eta}_{2\mu})
+ A_4(P_{\mu}\widetilde{\eta}_{1\nu} +P_{\nu}\widetilde{\eta}_{1\mu})
+ A_5(P_{\mu}\widetilde{\eta}_{2\nu} +P_{\nu}\widetilde{\eta}_{2\mu}),
\label{eq:eqW12}
\ee
where
$$\widetilde{g}_{\mu\nu}= g_{\mu\nu}-\frac{k_{\mu}k_{\nu}}{k^2},~
\widetilde{\eta}_{i\mu}=\eta_{i\mu}-\frac{k\cdot \eta_i}{k^2}k_{\mu},~i=1,2,
$$
and  
\ba 
A_1&=& \frac{G_M ^2}{2}(2k\cdot \eta_1 k\cdot \eta_2 -k^2\eta_1 \cdot \eta_2),~
A_2=-\eta_1 \cdot \eta_2 \frac{2M^2}{1+\tau} ( G_E^2(k^2)+\tau G_M^2(k^2)),
\nn\\
A_3&=& G_M^2(k^2)\frac{k^2}{2},~
A_4=-MG_M(k^2)\frac{ G_E(k^2)+\tau G_M(k^2)}{1+\tau}k\cdot \eta_2,~
\nn\\
A_5&=&MG_M(k^2)\frac{ G_E(k^2)+\tau G_M(k^2)}{1+\tau}k\cdot \eta_1.~\nn
\ea
The dependence of the polarization of the scattered proton on the polarization state of the proton beam is obtained by contraction of the spin independent leptonic tensor, not averaged over the spin of the initial electron, i.e., $2L_{\mu\nu}^{(0)}$ [Eq. (\ref{eq:eqLp})], and the spin-dependent hadronic tensor $W_{\mu\nu}(\eta_1, \eta_2)$ [Eq. (\ref{eq:eqW12})].

One obtains the following formula which holds in any reference system:
\ba
{\cal D}D(\eta_1, \eta_2)&=&2(1+\tau
)^{-1}\Bigl\{k\cdot\eta_1k\cdot\eta_2G_M(k^2)\left[k^2\left(G_M(k^2)-G_E(k^2)\right)+2m^2(1+\tau
)G_M(k^2)\right]\nn\\
&&
+k^2(1+\tau)G^2_M(k^2)(2k_1\cdot\eta_2k_2\cdot\eta_1-m^2\eta_1\cdot\eta_2)
\nn\\
&&
+4G_M(k^2)
(k\cdot\eta_1k_1\cdot\eta_2-k\cdot\eta_2k_1\cdot\eta_1)\left [M^2\tau
\left (G_E(k^2) -G_M(k^2)\right)\right .\nn\\
&&\left .
+mE\left(G_E(k^2)+\tau G_M(k^2)\right )\right ]
\nn\\
&&-\eta_1\cdot\eta_2\left(G^2_E(k^2)+\tau G^2_M(k^2)\right )
\left [k^2(M^2-2mE)+4m^2E^2\right ]\Bigr\}. 
\label{eq:eq29}
\ea

Applying the $P$-invariance of the hadron electromagnetic interaction, one can write the following expression for the dependence of the differential cross section on the polarization of the incident and scattered protons which participate in the reaction:
\be
\frac{d\sigma}{dk^2}(\eta_1,\eta_2)= \left (\frac{d\sigma}{dk^2}\right )_{un}
\left [1+
D_{tt}S_{1t}S_{2t} + 
D_{nn} S_{1n} S_{2n}+
D_{\ell\ell} S_{1\ell} S_{2\ell}+
D_{t \ell} S_{1t} S_{2\ell}+ 
D_{\ell t} S_{1\ell}S_{2t}\right ],
\label{eq:eqpol3}
\ee
where $D_{ik}$, $i,k=\ell,t,n$ are the corresponding spin depolarization coefficients which characterize $\vec p+e\to \vec p+ e$ scattering. 
The explicit expressions of the depolarization coefficients are given in Appendix \ref{Poldep} in terms of the hadron form factors.

\section{Numerical results}
\subsection{Experimental observables }

For a given proton-beam energy $E$ the observables are functions of only one kinematical variable, which we chose as $k^2$, because it is a kinematical invariant. Transformation to the scattering electron angle are straightforward.
The proton structure is taken into account through the parametrization of FFs.
We took the dipole parametrization
\be
G_E(k^2)=G_M(k^2)/\mu_p=[1-k^2/0.71]^{-2}, 
\label{eq:eqdipole}
\ee
where $\mu_p$ is the proton magnetic moment and $k^2$ is expressed in GeV$^2$. The normalization to the static point is $G_E(0)=1$ and $G_M(0)=\mu_p$. The standard dipole parametrization coincides with more recent descriptions for $-k^2<1$ GeV$^2$. At higher $k^2$,  different choices may affect the cross section and, to a lesser extent, the polarization observables. However as we showed above, the maximum value of $k^2$ which can be achieved in inverse kinematics, justifies the choice of dipole parametrization, and even of constant FFs, where the constants correspond to the static values.

\begin{figure}
\mbox{\epsfxsize=12.cm\leavevmode \epsffile{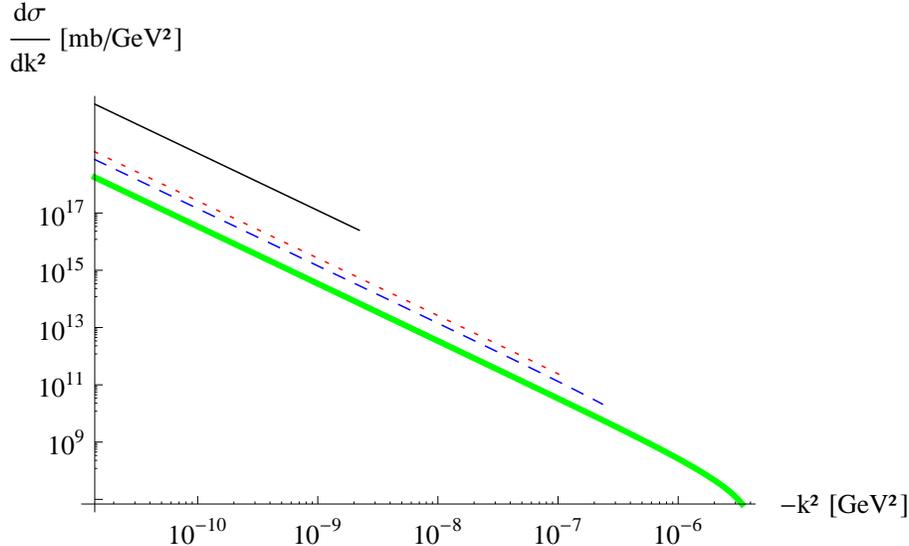}}
\vspace*{.2 truecm}
\caption{(Color online) Differential cross section as a function of $-k^2$ for different incident  energies: $E=1$ MeV (black solid
line), $E=50$ MeV (red dotted line), $E=100$ MeV (blue dashed line), $E=1$ GeV (thick green line).}
\label{Fig:figsia}
\end{figure}

\begin{figure}
\mbox{\epsfxsize=12.cm\leavevmode \epsffile{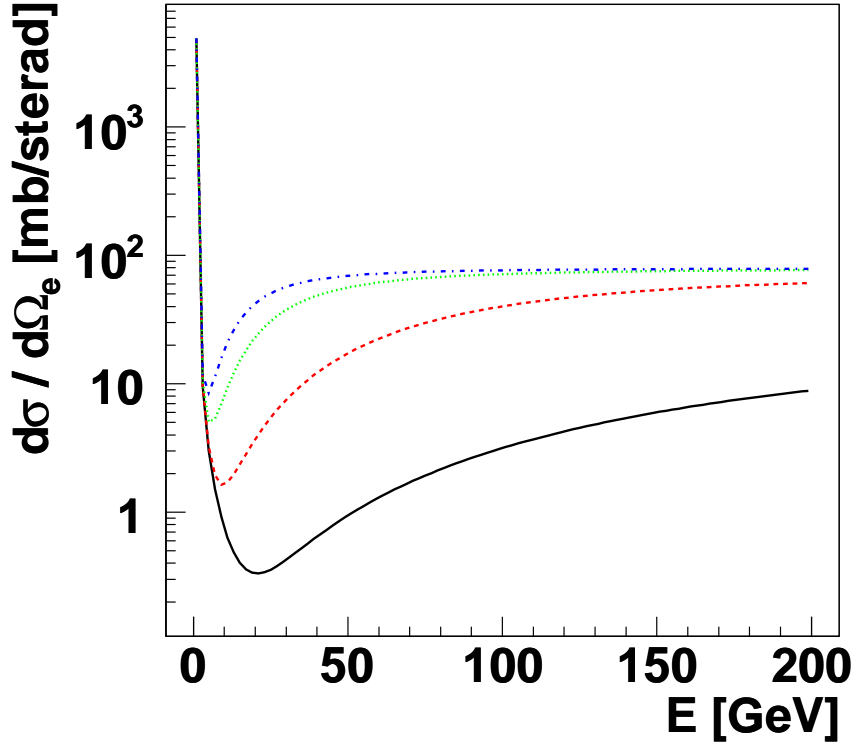}}
\vspace*{.2 truecm}
\caption{(Color online) Differential cross section as a function of incident  energy $E$ for different angles: $\theta_e=0$ (black solid line), 10 mrad
(red dashed line), 30 mrad (green dotted line), 50 mrad (blue dash-dotted line).}
\label{Fig:figsen}
\end{figure}

The differential cross section [Eq. (\ref{eq:eqSk})] is plotted as a function of $(-k)^2$ in Fig. \ref{Fig:figsia}. One can see that it is monotonically decreasing as a function of $k^2$ up to a value of $k^2_{max}$ according to Eq. (\ref{eq:kmax}).
\begin{figure}
\mbox{\epsfxsize=10.cm\leavevmode \epsffile{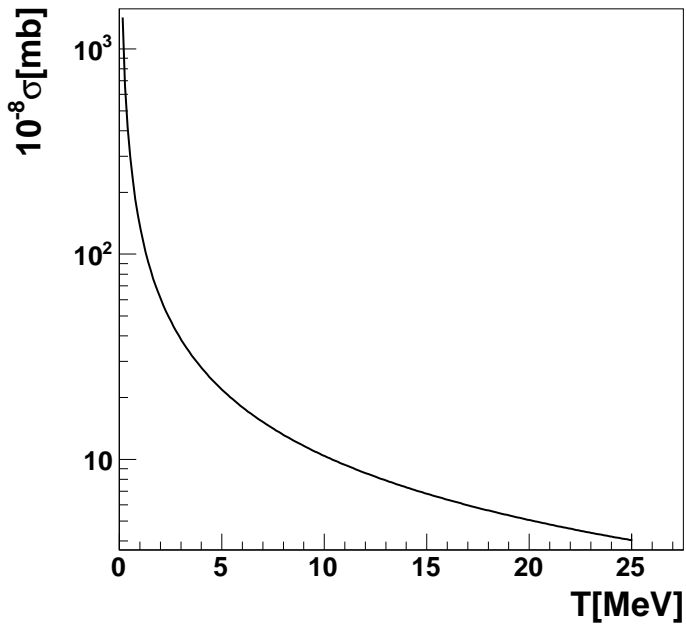}}
\vspace*{.2 truecm}
\caption{Total unpolarized cross section  as a function of incident proton kinetic energy $T$.}
\label{Fig:figsigm}
\end{figure}

The energy dependence of the cross section for different angles: $\theta_e=0$ (black solid line), 10 mrad
(red dashed line), 30 mrad (green dotted line), 50 mrad (blue dash-dotted line) is given in Fig. \ref{Fig:figsen}. The unpolarized differential cross section is divergent at small values of energy; it has an angle dependent minimum and then increases smoothly up to large energies.

As shown in Sec. II, Eq. (\ref{eq:eqSk}), the cross section diverges for $k^2\to 0$. This condition is obtained when the scattering angle is small (high energies, and large impact parameters), or when the energy is small.

In the first case, one introduces a minimum scattering angle, which is related to the impact parameter, whose classical $c$ and  $q$ quantum expressions are given by \cite{Ja75}:
\be
\theta_{min}^{(c)}=\displaystyle\frac{2 e^2}{p\beta b},
~\theta_{min}^{(q)}=\displaystyle\frac{\hbar}{p b}, 
\label{eq:thmin}
\ee
where $b$ is the impact parameter and $\beta$ is the relative velocity. Let us  take as the characteristic impact parameter, the Bohr radius, $b=0.519\times 10^{5}$ fm. We have shown above that there is a maximum scattering angle for the proton, which does not depend on the energy, and a corresponding maximum value for the transferred momentum $k^2$. The condition $k_{min}<k_{max}$ from Eqs. (\ref{eq:thmin}) is obtained for $E\ge 1$ MeV.
When the relative energy is very low, the electron and proton may be trapped in a bound system, and elastic scattering based on one-photon exchange cannot be applied to this process. The Born approximation corresponds to the first  term of an expansion in the parameter $\alpha/v$ which should be less than unity. The condition
$\alpha/v=0.1c$ is satisfied for  $E>2.5$ MeV. 

The description of Coulomb effects at low energies require approximations and is outside the purpose of this paper. We apply the present calculation for $E\ge$ 3 MeV. Screening effects may be important at low energies. They are introduced multiplying the cross section by the factor
\be
\chi= \displaystyle\frac{\chi_b}{e^{\chi_b}-1},~
\chi_b=-2\pi\displaystyle\frac{\alpha}{\beta}.
\label{eq:chi}
\ee
Such a factor is attractive for opposite charges and increases the cross section for the reaction of interest here. We apply this factor in our calculation. 
At the lowest energy ($E=$ 3 MeV) this correction is of the order of 30\%.
 
The total cross section has been calculated by integration from a value of $k^2_{min}$ extracted from Eqs. (\ref{eq:k2eps2}), (\ref{eq:eqep2}), and (\ref{eq:thmin}), and it is given as a function of incident proton kinetic energy $T=E-M$ in Fig. \ref{Fig:figsigm}, for values of $T$ in the MeV range.

\begin{figure}
\mbox{\epsfxsize=15.cm\leavevmode \epsffile{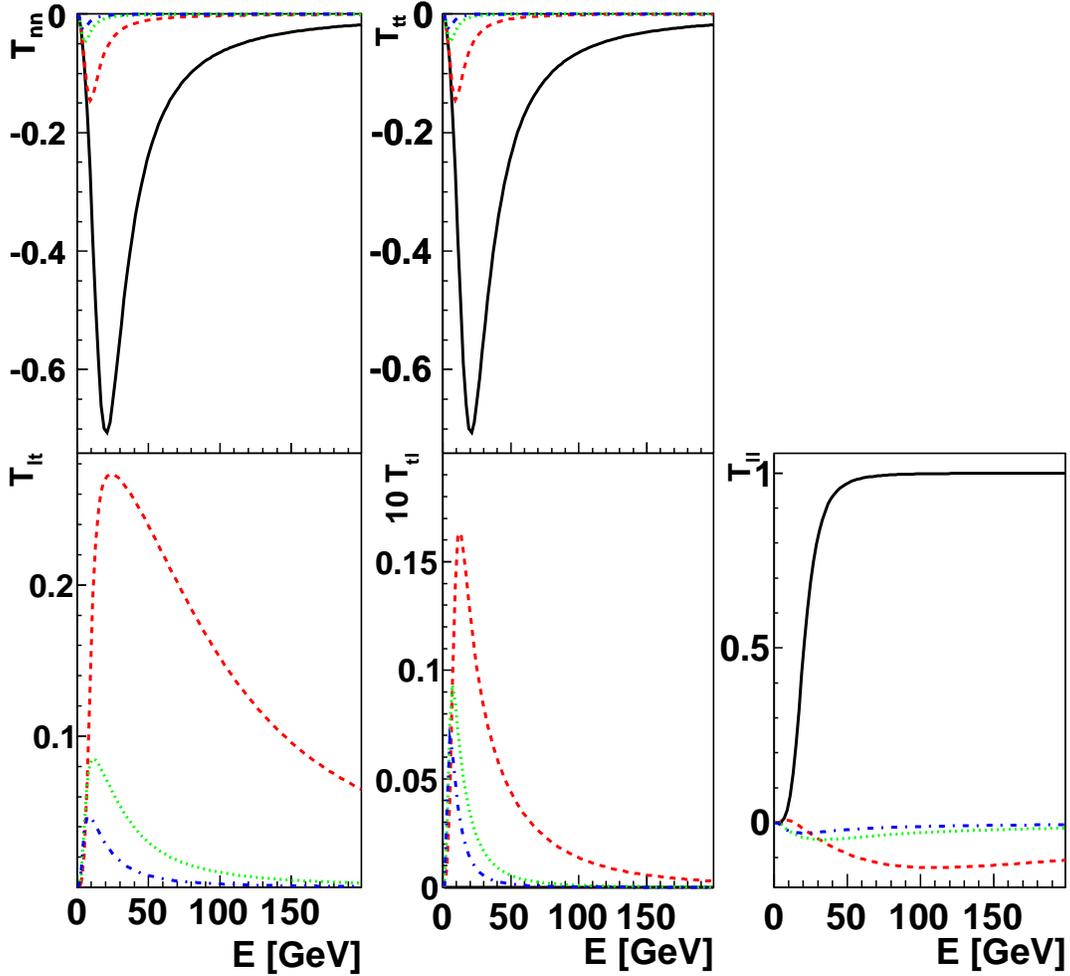}}
\vspace*{.2 truecm}
\caption{(Color online) Polarization transfer coefficients as a function of $E$ for different angles. Notations are the same as in Fig. \protect\ref{Fig:figsen}.}
\label{Fig:figtt}
\end{figure}

The polarization transfer coefficients [Eq. (\ref{eq:eqT})] are shown in Fig. \ref{Fig:figtt} as a function of the incident energy for $\theta=0$ (black solid line), 10 mrad (red dashed line), 30 mrad (green dash-dotted line), 50 mrad (blue dotted line).

\begin{figure}
\mbox{\epsfxsize=15.cm\leavevmode \epsffile{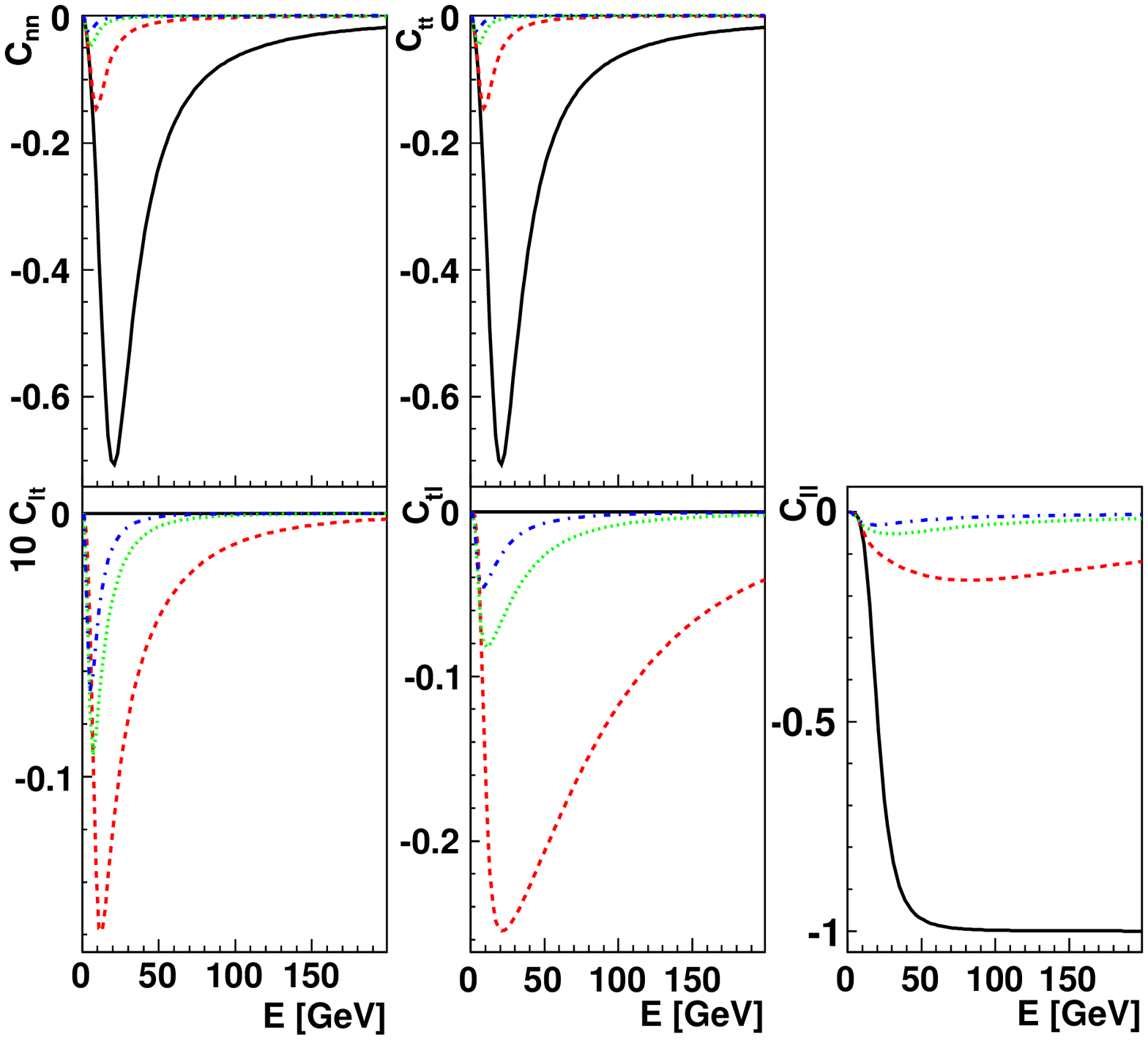}}
\vspace*{.2 truecm}
\caption{(Color online) Same as Fig. \protect\ref{Fig:figtt} but for the spin correlation coefficients.}
\label{Fig:figaa}
\end{figure}

\begin{figure}
\mbox{\epsfxsize=15.cm\leavevmode \epsffile{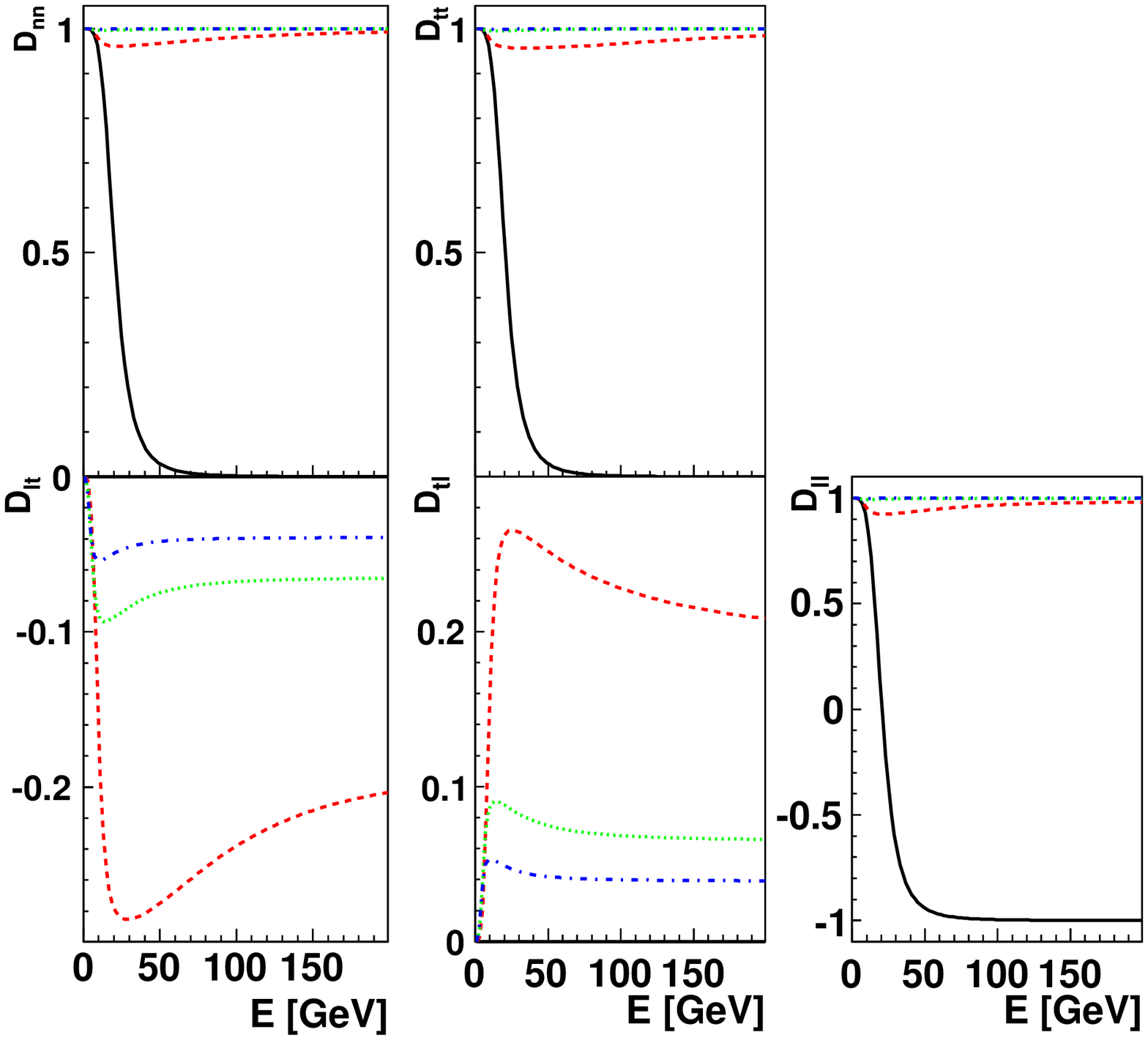}}
\vspace*{.2 truecm}
\caption{(Color online) Same as Fig. \protect\ref{Fig:figtt} but for the spin depolarization coefficients.}
\label{Fig:figdd}
\end{figure}

The spin correlation  coefficients [Eq. (\ref{eq:eqA})] are shown in Fig. \ref{Fig:figaa}. The spin depolarization 
coefficients, [Eq. (\ref{eq:eqD})], are shown in Fig. \ref{Fig:figdd}.

One can see that in collinear kinematics, in general, either polarization 
observables take the maximal values or they vanish. An interesting kinematic region appears at $E=20$ GeV, where a structure is present in agreement with the results of Ref. \cite{Gl97}. 

Let us calculate the cross section for a non polarized proton beam colliding with a  polarized target:
\be
\sigma_{ij}=  \int \! N{\cal D}T_{ij}P_iP_j \, \mathrm{d}k^{2},~N=\frac{\pi\alpha^2}{2m^2p^2k^4}.
\label{eq:eqpol4}
\ee
Assuming $P_i=P_j=1$, the values for different incident energies
are reported in Table \ref{table:t1} 
for the total polarized and unpolarized cross sections 
and in  Table \ref{table:t2} for the corresponding integrated polarization coefficients. 

\begin{center}
\begin{table}[ht]
\scriptsize
    \begin{tabular}{ | c | c | c | c | c | c | c | }
    \hline
$T$ & $\sigma_{unp}$ & $\sigma_{t\ell}$  & $\sigma_{\ell t}$ & $\sigma_{\ell\ell}$  & $\sigma_{tt}$ & $\sigma_{nn}$ \\ 
$ [GeV]$  &[mb]  &[mb]&[mb] &[mb]&[mb] & [mb] \\ 
 \hline
$23 \times 10^{-3} $ & $4.4 \times 10^{8} $      &  $26 $    & $26.7 $    &$-125.3 $     &   $-16.9 $  &$-139.3 $     \\ 
\hline
  $50 \times 10^{-3} $& $2 \times 10^{8} $ & $11.5 $ & $12.2$ &$-62.8$ & $-7.4 $ &$-67$  \\ 
\hline
 $1 $ & $2.5 \times 10^{7} $ &0.4 & 0.8&  -5.6  & -0.2  & -2.9 \\ \hline
  $10 $ &  $1.9 \times 10^{7} $ &$9.1 \times 10^{-3}$  &$10.6 \times 10^{-2}$ & -1.01  & $-0.6 \times 10^{-2}$  & -0.09 \\ \hline 
  $50$ &$ 1.8 \times 10^{7} $ &$0.4 \times 10^{-3}$  &$2.3 \times 10^{-2}$ &-0.2    &  $-0.3 \times 10^{-3}$&$-0.5 \times 10^{-2}$    \\
\hline 
\end{tabular}
\caption{Unpolarized cross section and polarized transfer cross sections (in mb) for different incident energies.}
\label{table:t1}
\end{table}
\end{center}

\begin{center}
\begin{table}[ht]
\scriptsize
    \begin{tabular}{ |  c | c | c | c | c | c |}
    \hline
$T$ & $T_{t\ell}$ & $T_{\ell t}$ & $T_{\ell\ell}$ &  $T_{tt}$  & $T_{nn}$  \\ 
 $[GeV]$&  &  &  &  &  
\\ \hline
  $23\times 10^{-3} $  & $1.5\times 10^{-12} $ &$1.5\times 10^{-12} $ &$-1.3\times 10^{-12} $ &$-2.6\times 10^{-12} $ &$-3.8\times 10^{-12} $
  \\ \hline
  $50 \times 10^{-3} $ & $7.2 \times 10^{-12} $ &  $7.5 \times 10^{-12} $ & $-6.3 \times 10^{-12} $   & $-1.2 \times 10^{-11} $  & $-1.8 \times 10^{-11} $  \\ 
  \hline
  $1$  & $3.3\times 10^{-9}$  & $6.8\times 10^{-9}$& $-4.8\times 10^{-9}$ & $-6.8 \times 10^{-9}$ & $-9.2\times 10^{-9}$  \\ 
  \hline
  $10 $ & $3.5\times 10^{-7}$ & $3.9\times 10^{-6}$ & $-1.4\times 10^{-6}$ &  $-1.1 \times 10^{-6}$  & $-1.2\times 10^{-6}$  \\  
  \hline
  $50$ & $5.9\times 10^{-6} $ &$0.3\times 10^{-3} $   &$1.4\times 10^{-3} $  &$-1.4\times 10^{-5}$    &$-0.2\times 10^{-4} $   \\
   \hline
\end{tabular}
\caption{Integrated polarization coefficients for different incident energies.}
\label{table:t2}
\end{table}
\end{center}

\begin{figure}
\mbox{\epsfxsize=12.cm\leavevmode \epsffile{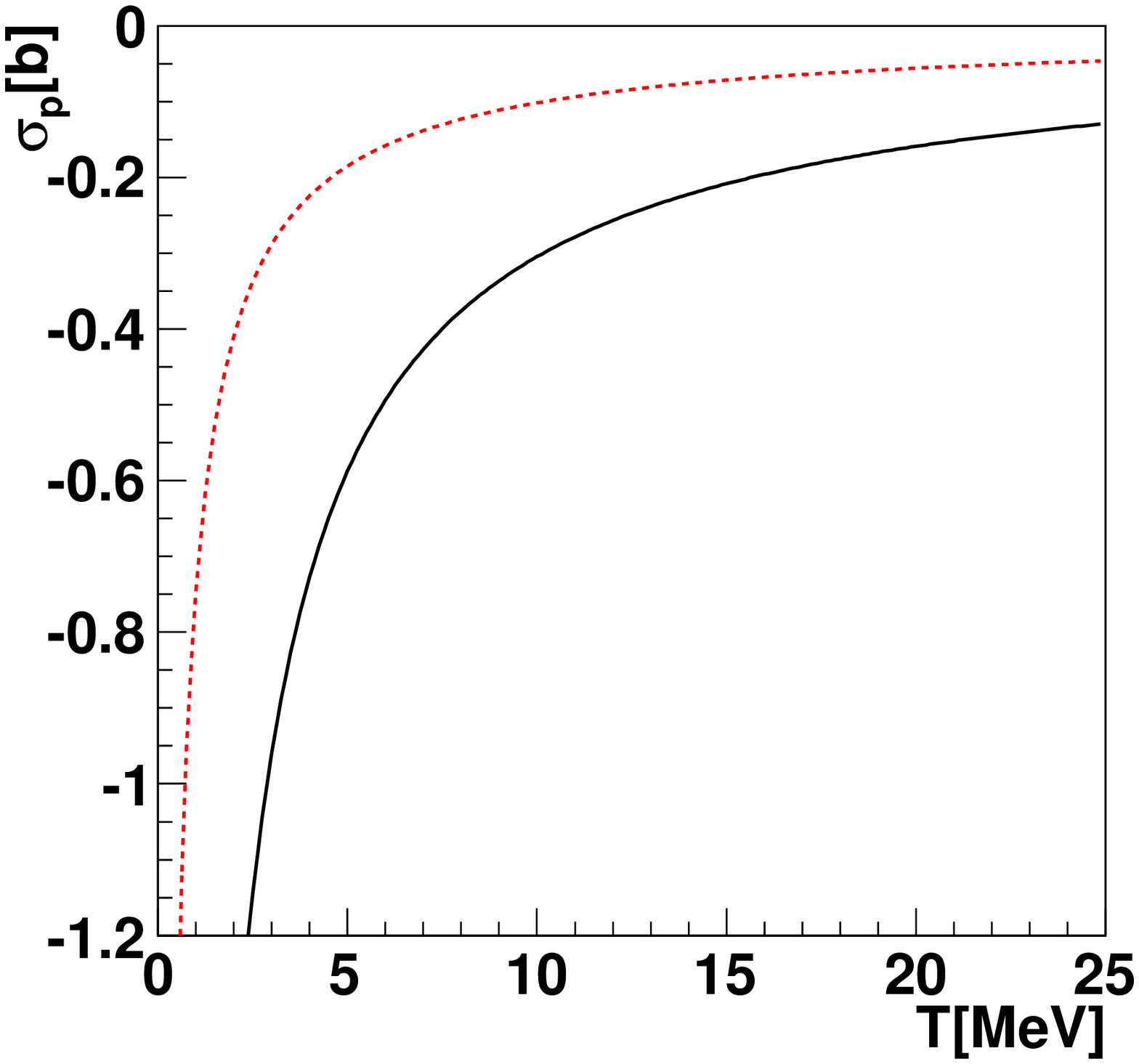}}
\vspace*{.2 truecm}
\caption{(Color online) Spin transfer cross section 
$\sigma_{\ell}=(\sigma_{\ell\ell}+\sigma_{\ell t})/2$ (black  solid line) and 
$\sigma_{nn}$ (red dashed line) as a function of proton kinetic energy $T$.}
\label{Fig:figsinn}
\end{figure}

The spin transfer cross section $\sigma_{nn}$ and $\sigma_{\ell}=(\sigma_{\ell\ell}+\sigma_{\ell t})/2$ are illustrated in Fig. \ref{Fig:figsinn} in the MeV range.

These values are very sensitive to the incident energy, and they are consistent with the findings of Refs. \cite{Ra93,Ho94,Gl96}. Although they cannot be compared directly with the previous calculations, because our formalism is derived in the laboratory system, they allow a more direct comparison to experiment.

\subsection{High energy polarimetry}
From Figs. \ref{Fig:figtt}, \ref{Fig:figaa} and  \ref{Fig:figdd}  it appears that polarization coefficients are in general quite large, except at low energy.  Proton electron scattering can be used, in principle, to measure the polarization of high-energy beams. The idea to use $pe$ elastic scattering for beam polarimetry has already been suggested in Refs. \cite{Gl96}. Let us calculate the figure of merit, for measuring the polarization of a secondary proton beam, 
after scattering from atomic electrons. 

\begin{figure}
\mbox{\epsfxsize=16.cm\leavevmode \epsffile{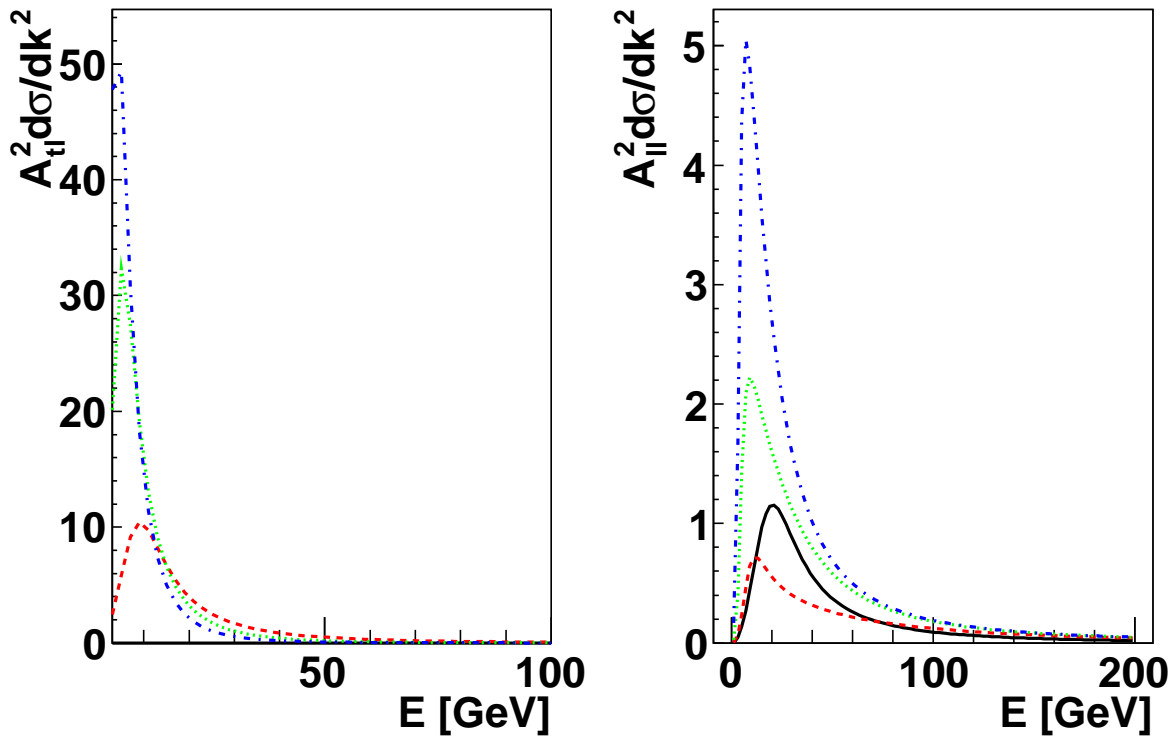}}
\vspace*{.2 truecm}
\caption{(Color online) Variation of differential quantities $ A_{t\ell}^2(k^{2})(d\sigma/dk^{2}) $ (left) and $ A_{\ell\ell}^2(k^{2})(d\sigma/dk^{2})$ (right) [a.u.] as a function of incident energy for a polarized proton beam scattering from a polarized electron target $\vec p+\vec e\to p+e$, at different angles.
Notations are the same as in Fig. \protect\ref{Fig:figsen}.
 }
\label{Fig:fom33}
\end{figure}

The differential figure of merit is defined as
$${\cal F}^2(\theta_p)=\epsilon(\theta_p)A^2_{ij}(\theta_p),$$
where $ A_{ij}$ stands for a generic polarization coefficient and $\epsilon(\theta_p)=N_f(\theta_p)/N_i$ is the number of useful events over the number of the incident events in an interval $\Delta\theta_p$ around $\theta_p$. Because it is related to the inverse of the statistical error on the polarization measurement, this quantity is useful for a proton with degree of polarization $P$:
\be
\left(\frac{\Delta P(\theta_p)}{P}\right )^2=\frac{2}{N_i(\theta_p){\cal F}^2(\theta_p)P^2}= \frac{2}{L t_{m}(d\sigma/d\Omega )d\Omega A_{ij}^2(\theta_p) P^2},
\label{eq:deltaP}
\ee
where $t_{m}$ is the time of measurement. 
The correlation coefficient squared, weighted by the differential cross section, $A_{t\ell}^2(k^{2})(d\sigma/dk^{2}) $ and $ A_{\ell\ell}^2(k^{2})(d\sigma/dk^{2})$ are shown in Fig. \ref{Fig:fom33} for  different electron angles.

\begin{figure}
\mbox{\epsfxsize=12.cm\leavevmode \epsffile{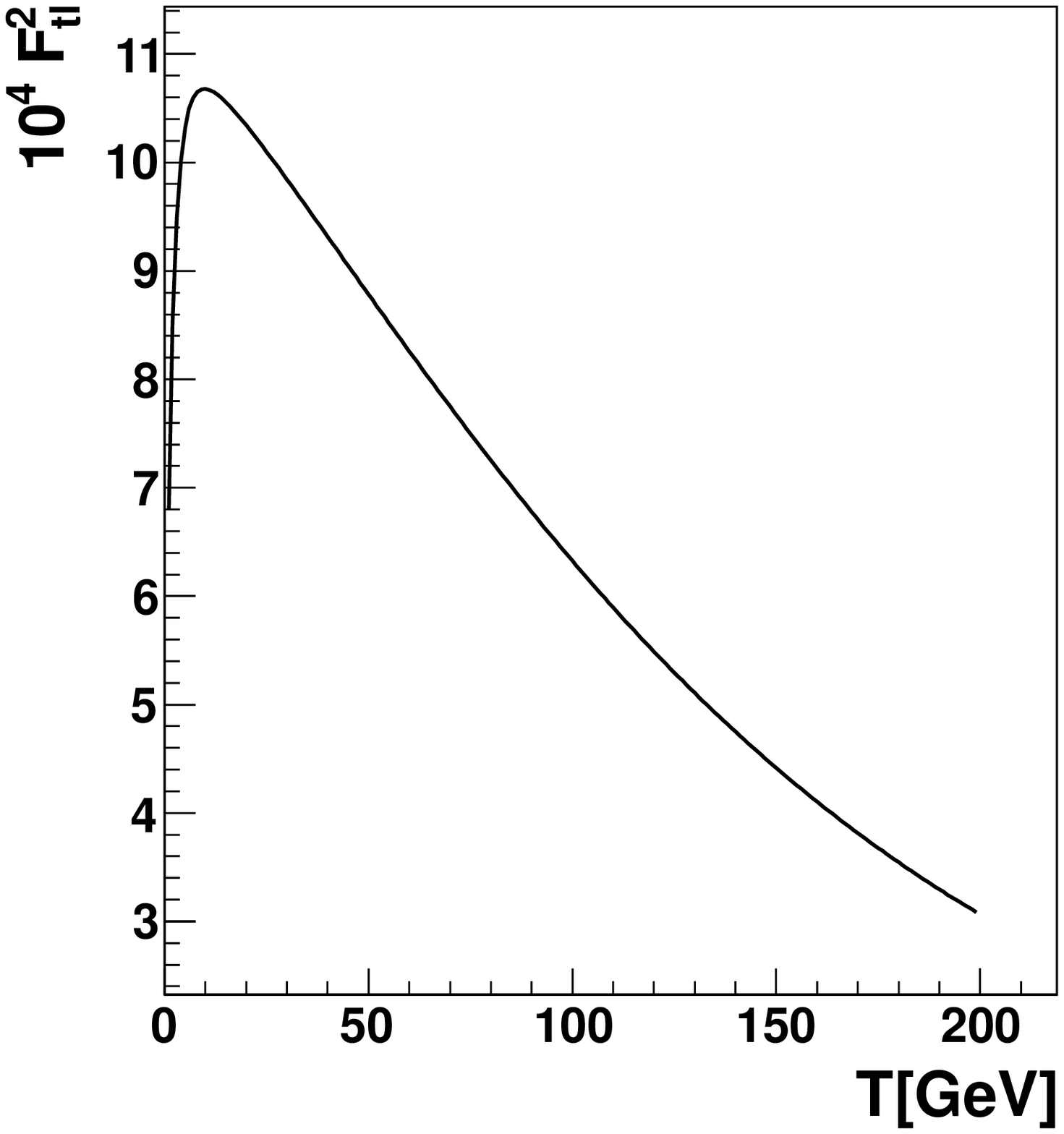}}
\vspace*{.2 truecm}
\caption{Variation of the quantity $F^{2}$ [a.u.] as a function of proton kinetic energy $T$ for a transverse polarized proton beam scattering from a longitudinally polarized electron target ($\vec p+\vec e\to p+e$).}
\label{Fig:fom22}
\end{figure}

The integrated quantity
\be
F^{2}=\int \!  \,\frac{d\sigma}{dk^{2} } A_{ij}^2(k^{2}) \mathrm{d}k^{2}
\label{eq:fm2}
\ee
as a function of the incident energy is shown in Fig. \ref{Fig:fom22}.
In Refs. \cite{Nikolenko} it was suggested to use the scattering of a transverse polarized proton beam from a longitudinally polarized electron target. From Fig. \ref{Fig:fom22}, one can see that $F^2$ takes its maximum value for $T\simeq 10$ GeV. Assuming a luminosity of $10^{32}$ cm$^{-2}$ s$^{-1}$ for an ideal detector with an acceptance and efficiency of 100\%, 
one could measure the beam polarization with an error of 1\% in a time interval of 3 min. 

If one detects the outgoing proton, which seems more challenging because its kinematical characteristics are close to those of the beam (for the high-energy solution) one could in principle build a polarimeter based on the scattering from the polarized beam (the polarization of which should be known) on an unpolarized target. In this case, from the azimuthal distribution, one can reconstruct the components of the polarization which are normal to the scattering plane.

\section{Conclusions}

The elastic scattering of protons from electrons at rest was investigated in a
relativistic approach in the one-photon-exchange (Born) approximation. This
reaction, where the target is three orders of magnitude lighter than the projectile,
has specific kinematical features due to the "inverse kinematics" (i.e., the
projectile is heavier than the target). For example, the proton is scattered at very
small angles and the allowed momentum transfer does not exceed the MeV$^2$ scale, even when the proton incident energy is of the order of GeV.  
The differential cross section and various double spin polarization observables have
been calculated in terms of the nucleon electromagnetic FFs. Note that, for the values
of transferred momentum involved, any parametrization of FFs is equivalent and is
very near to the static values. The spin transfer coefficients to a polarized
scattered proton were calculated for two cases: when the proton beam is polarized or when the electron target is polarized. The correlation spin coefficients when the proton
beam and the electron target are both polarized were also calculated. Numerical estimates showed that polarization effects may be sizable in the GeV
range, and that the polarization transfer coefficients for $\vec p+e\to \vec p+e$
could be used to measure the polarization of high energy proton beams. This result
confirms previous estimates from \cite{Gl96}.
The calculated values of the scattered proton polarization for the reaction $p+\vec
e\to \vec p+e$ at proton-beam energies lower then a few tens of MeV show that
it is not possible to obtain sizable polarization of the antiproton beam in an
experimental setup where antiprotons and electrons collide with small relative
velocities. The present results confirm that the polarization of the scattered proton
has large values at high proton-beam energies. Thus, one
could consider an experimental setup where high-energy protons collide with a
polarized electron target at rest. The low values of momentum transfer which are
involved ensure that the cross section is sizable.

\section{Acknowledgments}
One of us (A.D.) acknowledges the Libanese CNRS for financial support. This work was partly supported by  CNRS-IN2P3 (France) and by the National Academy of Sciences of Ukraine under PICS No. 5419 and by GDR No.3034 "Physique du Nucl\'eon" (France). This work was initiated in collaboration with M. P. Rekalo. We acknowlege E.A. Kuraev, D. Nikolenko, J. Van de Wiele, P. Lenisa and F. Rathmann for interesting discussions on different aspects related to electron elastic scattering and polarization phenomena. Xu Gang is thanked for participation in the early stage of this work. 
\appendix
\section{Appendix A: Polarization transfer coefficients, $T_{ij}$, for  $p+\vec e\to \vec p+e$}
\label{Poltra}
The explicit expressions for the polarization transfer coefficients for $p+\vec e\to \vec p+e$ are: 
\ba
{\cal D}T_{nn}&=&4mM k^2G_E(k^2) G_M(k^2),\nn\\
{\cal D}T_{tt}&=& 4mM k^2\frac{G_M(k^2)}{1+\tau}\left \{(1+\tau)G_E(k^2) - \left (E+M+\frac{k^2}{2m}\right )^{-1}  
\left(1-\frac{k^2}{k^2_{max}}\right) \right. \nn\\
&&
\left. \left[ (E+M+2E\tau)  G_E(k^2)- 
\tau (E+M+2M\tau) G_M(k^2) \right ] \right \}, \nn\\
{\cal D}T_{t\ell}&=&-2mpk^2\frac{G_M(k^2)}{1+\tau}
\left (E+M+\frac{k^2}{2m}\right )^{-1} 
\left [ -k^2\left (1- \frac{k^2}{k^2_{max}}\right )\right ]^{1/2}\nn\\
&&
\left \{ \frac{M}{m}\frac{m+M}{E-M} 
\left [(1+2\tau)G_E(k^2)- 
\tau G_M(k^2)\right ]+\right .\nn\\
&&
\left.
\left (1-2m\frac{E+m}{s}\frac{k^2}{k^2_{max}}\right )\left [G_E(k^2)+
\tau G_M(k^2)\right ] \right \}, \nn\\
{\cal D}T_{\ell t}&=&-4mp k^2\frac{G_M(k^2)}{1+\tau}
\left (E+M+\frac{k^2}{2m}\right )^{-1} 
\left[-k^2\left (1- \frac{k^2}{k^2_{max}}\right )\right ]^{1/2}\nn\\
&&
\left\{ (1+\tau)G_M(k^2)+\frac{E-M}{2M}\left[  G_M(k^2)-G_E(k^2)\right ]
\right .\nn\\
&&
- m\frac{E+m}{s} \frac{1}{k^2_{max}}
\left [ k^2 \left (G_E(k^2)+\tau G_M(k^2)\right )+ 
\right .\nn\\
&&
\left.\left.
2M(E+M)\left ( G_E(k^2)(1+2\tau) -\tau G_M(k^2)\right )\right ] \right \},\nn\\
{\cal D}T_{\ell \ell}&=&4mM k^2\frac{G_M(k^2)}{1+\tau}
\left \{(1+\tau)
\left[ \frac{E}{M}+ \frac{xk^2}{2m}(E+M+2m) -\right .\right .\nn\\
&&
\left.\frac{(E+m)^2}{s}\frac{k^2}{k^2_{max}} \left ( 1+ \frac{xk^2}{2m}
(m-M)\right )+
\frac{1}{s}(m+M)(E+m)xp^2 \frac{k^2}{k^2_{max}}\right ] G_E(k^2)\nn\\
&&
+\tau \left [ xp^2 \frac{M+m}{m}\left (1-3 \frac{m(E+m)}{s} \frac{k^2}{k^2_{max} }\right )+ \right. 
\nn\\
&&\left .\left .
\frac{(E+m)^2}{s}\frac{k^2}{k^2_{max}} 
\left ( 1-\frac{xk^2}{2m}(m+M)\right ) - \frac{E+m}{m}\right]
\left [G_M(k^2)- G_E(k^2)\right ]\right \},
\label{eq:eqT}
\ea
with $x^{-1}=M(E+M+\frac{k^2}{2m})$, and $s=m^2+M^2+2mE$ is the total energy in the proton electron elastic scattering.
\section{Appendix B: Polarization correlation coefficients, $C_{ij}$, for $\vec  p+\vec e \to p+e $}
\label{Polcor}
The explicit expressions of the spin correlation coefficients, as a function of the Sachs FFs can be written as:
\ba
{\cal D} C _{nn}&=&4mM k^2G_E(k^2)G_M(k^2),\nn\\
{\cal D}C_{tt}&=& 4mM \tau k^2\frac{G_M(k^2)}{1+\tau}
\left[(1-\frac{4M^2}{k^2_{max}})G_E(k^2)
+(\frac{k^2}{k^2_{max}}-1)G_M(k^2)\right],\nn\\
{\cal D}C_{t \ell}&=&8mMp \left [-k^2  \left (1-\frac{k^2}{k^2_{max}} 
\right )\right ]^{1/2}\frac{G_M(k^2)}{1+\tau}
\Big \{\tau \left [G_M(k^2)- G_E(k^2)\right ]
\nn\\
&&
 \left .-\frac{k^2}{k^2_{max}}\frac{m(E+m)}{s} 
 \left [\tau G_M(k^2)+ G_E(k^2)\right ]
\right \},
\nn\\
{\cal D} C_{\ell t}&=&-2mM \frac{k^2}{p}
\left ( \frac{E}{M}-\frac{M}{m} \right ) 
\left [-k^2\left ( 1-\frac{k^2}{k^2_{max}}\right )\right]^{1/2} 
\frac{G_M(k^2)}{1+\tau}\left [\tau G_M(k^2)+ G_E(k^2)\right ],\nn\\
{\cal D} C_{\ell\ell }&=&4 k^2\frac{G_M(k^2)}{1+\tau}
\left\{ (mE-\tau M^2) G_E(k^2)+\tau(M^2+mE) G_M(k^2)
\right .
\nn\\
&&\left.
-(M^2+mE)\frac{k^2}{k^2_{max}}\frac{m(E+m)}{s} 
\left [\tau G_M(k^2)+ G_E(k^2)\right ]\right \}.
\label{eq:eqA}
\ea
\section{Appendix C: Depolarization coefficients, $D_{ij}$, for $\vec p+ e \to \vec p+e $}
\label{Poldep}
The depolarization coefficients from the polarized beam to the ejectile for $\vec  p+ e \to \vec p+e $, are expressed in terms of the hadron form factors, as:
\ba
{\cal D} D_{tt}&=&-R_1 -k^2\left(1-\frac{k^2}{k^2_{max}}\right)\left \{ \frac{m}{M}
\left (R_3-R_4\right ) - xR_1+\left ( 1-xk^2\frac{m+M}{2m}\right )R_2\right \},\nn\\
{\cal D} D_{nn}&=&-R_1,\nn\\
{\cal D} D_{\ell\ell}&=&\frac{R_1}{M}\left \{\frac{p}{M}\left (p+\frac{k^2}{2m}\frac{E+m}{p}\right )-E\left [1+x\left (p+\frac{k^2}{2m}\frac{E+m}{p}\right )^2 \right ]\right \}+ \nn\\
&& 
R_4\frac{m}{M}\frac{k^2}{2m}\left \{
\frac{1}{M}\left (p-E\frac{E+m}{p}\right )
\left (p+\frac{k^2}{2m}\frac{E+m}{p}\right )+E+m-
\right .\nn\\
&& 
\left .p\left (p+\frac{k^2}{2m}\frac{E+m}{p}\right )
\left [ \frac{1}{M}-x\left (E-m+\frac{k^2}{2m}\right )\right ]
\right \}+
R_2 \frac{1}{M}\left (\frac{k^2}{2m}\right )^2
\left ( p-E\frac{E+m}{p}\right )
\nn\\
&& \left \{\left ( p+\frac{k^2}{2m}\frac{E+m}{p}\right )
\left [ \frac{1}{M}-x\left(E-m+\frac{k^2}{2m}\right )\right ]
-\frac{E+m}{p} \right \}+
\nn\\
&& 
R_3 \frac{m}{M}p \left \{\frac{k^2}{2m}\frac{E+m}{p}+
\left ( p+ \frac{k^2}{2m} \frac{E+m}{p} \right )
\left [  \frac{m}{M} - \frac{1}{M} \frac{k^2}{2m} +x \frac{k^2}{2m}
\left ( E-m+ \frac{k^2}{2m}\right )\right ]\right \}+
\nn\\
&& 
R_3 \frac{m}{M^2}\left ( p+ \frac{k^2}{2m}\frac{E+m}{p}\right )
\left [ mp+ \frac{1}{p} \frac{k^2}{2m}(M^2+mE)\right ],
\nn\\
{\cal D} D_{t \ell}&=&\frac{1}{p}
\left [ -k^2\left (1- \frac{k^2}{k^2_{max}}\right )\right ]^{1/2}
\left \{ x\left [ p^2+\frac{k^2}{2m}(E+m)\right ]\right .
\nn\\
&&
\left .\left [R_1+\frac{k^2}{2m}(m+M)R_2+\frac{m}{Mx}(R_4-R_3) \right ]-\frac{k^2}{2m}(E+m)R_2 \right \}, 
\nn\\
{\cal D} D_{\ell t}&=&\frac{1}{M^2p}\left[-k^2\left (1- \frac{k^2}{k^2_{max}}\right )\right ]^{1/2}
\left\{ -mp^2 \left [ MR_4+(M+2m)R_3\right ]+\right .\nn\\
&&
\frac{k^2}{2}(M^2+mE)\left(R_4-R_3-\frac{M}{m}R_2\right )+\nn\\
&&
\frac{xk^2}{2m}M(m+M) \left [mp^2(R_3+R_4)+(M^2+mE) \frac{k^2}{2m}R_2 \right ]+
\nn\\
&&
\left . xMR_1
\left [\frac{k^2}{2m}(M^2+mE)-Mp^2\right ]\right \},
\label{eq:eqD}
\ea
where 
\ba
R_1&=&-2\left [m^2k^2G_M^2+\frac{G_E^2+ \tau G_M^2}{1+\tau}( M^2k^2 +2 mEk^2+4m^2E^2)\right ],
\nn\\
R_2&=& 2 \frac{ G_M}{1+\tau}\left [2m^2(1+\tau)G_M+k^2 (G_M-G_E)\right ],\nn\\
R_3&=&2k^2G_M^2,\nn\\
R_4&=& 2(k^2+4mE) G_M\frac{G_E+ \tau G_M}{1+\tau}.
\label{eq:eqR}
\ea


\end{document}